 \documentstyle[12pt]{article}
\begin{document}
 \newcommand{\be}[1]{\begin{equation}\label{#1}}
 \newcommand{\ee}{\end{equation}}
 \newcommand{\beqn}[1]{\begin{eqnarray}\label{#1}}
 \newcommand{\eeqn}{\end{eqnarray}}
\newcommand{\mat}[4]{\left(\begin{array}{cc}{#1}&{#2}\\{#3}&{#4}\end{array}
\right)}
 \newcommand{\matr}[9]{\left(\begin{array}{ccc}{#1}&{#2}&{#3}\\{#4}&{#5}&{#6}\\
{#7}&{#8}&{#9}\end{array}\right)}
 \newcommand{\eps}{\varepsilon}
 \newcommand{\Ga}{\Gamma}
 \newcommand{\la}{\lambda}
 \renewcommand{\thefootnote}{\fnsymbol{footnote}}
\begin{flushright}
INFN-FE-05-94 \\
June  1994
\end{flushright}
\vspace{4mm}

 \begin{center}
 {\large\bf New Predictive Framework for Fermion Masses } \\
\vspace{2mm}

{\large\bf  in SUSY SO(10)}\footnote{Talk given at the  `SUSY 94' Workshop,
Ann Arbor, USA, 14-17 May 1994. To appear in the Proceedings } \\

\vspace{5mm}
{\large  Zurab G. Berezhiani}\footnote{E-mail: BEREZHIANI@FERRARA.INFN.IT,
31801::BEREZHIANI}\\ [5mm]
{\em Istituto Nazionale di Fisica Nucleare, Sezione di Ferrara, 44100 Ferrara,
Italy, \\ [2mm]
and\\ [2mm]
Institute of Physics, Georgian Academy
of Sciences, 380077 Tbilisi, Georgia}\\ [8mm]
\end{center}

\vspace{2mm}
\begin{abstract}


We present a new predictive approach based on SUSY $SO(10)$ theory.
The inter-family hierarchy is first generated in the sector of hypothetical
superheavy fermions and then transfered inversely to ordinary
quarks and leptons by means of the universal seesaw mechanism.
The obtained mass matrices are
simply parametrized by two small complex coefficients $\eps_u$ and $\eps_d$,
which can be given by the ratio of the GUT scale $M_G\simeq 10^{16}$ GeV
and some higher scale $M\simeq 10^{17}-10^{18}$ GeV (presumably superstring
scale). The model provides a possibility for
doublet-triplet splitting without fine tuning and the Higgsino mediated $d=5$
operators for the proton decay are naturally suppressed.
Our ansatz provides the correct {\em qualitative} picture of fermion mass
hierarchy and mixing pattern, provided that $\eps_d/\eps_u\sim 10$.
The running masses of the first family fermions: electron, u-quark
and d-quark obey an approximate $SO(10)$ symmetry limit. At GUT scale we have:
$u\sim d\simeq 3e$, $(\frac{\eps_u}{\eps_d})c\sim s\simeq \frac{1}{3}\mu$ and
$(\frac{\eps_u}{\eps_d})^2t\sim b\simeq \tau$.
The Cabibbo angle is large: $s_{12}\simeq \sqrt{m_d/m_s}$
while other mixing angles have their natural size:
$s_{23}\sim m_s/m_b$ and $s_{13}\sim m_d/m_b$.
We have many strong {\em quantitative} predictions though no special `zero'
texture is utilized  (in contrast to the known predictive frameworks).
Namely, taking as input the lepton, c-quark and b-quark masses, $m_s/m_d$
mass ratio and Cabibbo angle, we can
obtain the light (u,d,s) quark masses, top mass and $\tan\beta$.
The top quark is naturally in the 100 GeV range, but not too heavy:
$m_t<165$ GeV. The lower bound $M_t>150$ GeV (160 GeV) implies
$m_s/m_d>19$ ($>22$). $\tan\beta$ can vary from 1.2 to 1.7.
For light quark masses we have $m_s\approx 150$ MeV, $m_d\approx 7$ MeV,
and $m_u/m_d\simeq 0.5$.

\end{abstract}
\renewcommand{\thefootnote}{\arabic{footnote})}
\setcounter{footnote}{0}

\newpage


{\bf 1. Family Puzzles and SUSY GUT}

\vspace{3mm}

Understanding the origin of the observed pattern of fermion masses and
mixing is one of the main issues in modern particle physics. The
Standard Model accomodates all the observed quarks (including Top) and
leptons in a consistent way --  three families bear the same quantum numbers
under the $G_{SM}=SU(3)_C\otimes SU(2)_L\otimes U(1)_Y$ gauge symmetry.
The fermion mass scale is set by the same Higgs field that gives
masses to $W^{\pm}$ and Z bosons.
However, the pattern of fermion masses and mixing remains undetermined
due to arbitrariness of the Yukawa couplings: there is no explanation, what
is the origin of hierarchy of their eigenvalues, why they are approximately
alligned for the up and down quarks thus providing the small mixing, what
is the origin of their
complex structure, why the $\Theta$-term is vanishing in spite of this
complex structure etc. Understanding of all these issues
calls for a more fundamental flavour theory beyond the Standard Model.

There is an almost holy trust that all the fundamental problems, and among
them the problem of fermion masses, will find a final solution within the
Superstring Theory -- `Theory of Everything'. In principle, it should allow us
to {\em calculate} all Yukawa couplings. Unfortunatelly, we do not know
how the superstring can be linked to lower energy physics
in unambiguous way. Many theorists try to attack the problem in whole,
or its certain aspects, in the context of particular (among many billions)
superstring inspired models. However, the problem remains far away from
being solved and all what we know at present from superstring can be updated
in few important but rather general recommendations.

Nevertheless, one may think that many aspects of the flavour problem can
be understood by means of more familiar symmetry properties. A relevant theory
could take place at some intermediate energies between the electroweak and
Planck scales. Nowadays the concepts of grand unification and supersymetry are
the most promising ideas towards the physics beyond the SM, providing a sound
basis for understanding the issues
of coupling unification and stability of gauge hierarchy. In particular, the
famous coupling crossing phenomenon in the Minimal Supersymmetric Standard
Model (MSSM)  points to the GUT scale $M_G\simeq 10^{16}$ GeV
\cite{Amaldi}.

The $SO(10)$ model is an outstanding candidate for GUT. It unifies all
quark and lepton states in one family into one irreducible representation 16,
providing thereby possibility to link their masses with specific
group-theoretical relations. This can be clearly seen by considering the
$G_{PS}=SU(4)\otimes SU(2)_L\otimes SU(2)_R\otimes P$
subgroup of $SO(10)$:
$SU(4) \supset SU(3)_C\otimes U(1)_{B-L}$ unifies the lepton with quarks
as a fourth colour, whereas the $SU(2)_L\otimes SU(2)_R$
provides an "isotopic" symmetry interchanging the
up and down fermions, so that their mass matrices are somehow alligned even
if this symmetry is spontaneously broken.
In this way the smallness of quark mixing angles can be naturally linked
to the quark mass 'horizontal' hierarchy, though the origin of this
hierarchy in itself is beyond the scope of the $SO(10)$ model. The automatic
discrete $ P(L\leftrightarrow R)$ symmetry, essentially P-parity,
can be also used for constraining the fermion mass matrices.
Therefore, the $SO(10)$ naturally
contains all types of the simplest fermionic symmetries: the isotopic and
quark-lepton symmetries as well as P-parity, which
appear to be necessary but not sufficient tools for the fermion
mass model building: in order to constrain the fermion mass matrices at the
needed degree, also some inter-family (horizontal) symmetry ${\cal H}$
should be invoked.
It is natural to expect that ${\cal H}$ is also broken at the GUT scale $M_G$.
Such a $SUSY\otimes SO(10)\otimes {\cal H}$ theory can be regarded as a
Grand Unification of fermion masses.

On the other hand, we believe that the Standard Model is literally correct at
lower energies. This sets a 'boundary' condition for any hypothetical theory of
the flavour: in the low energy limit it should reduce to the minimal Standard
Model (or, better to say to MSSM \cite{Kane}) in  all sectors
(gauge, fermion and Higgs) -- all the possible extra
degrees of freedom must decouple. Since the decoupling is expected to occur
at superhigh ($\sim M_G$) energies, there is practically no hope to observe
experimentally any direct dynamical effect of such a flavour theory.
However, it could manifest itself in the sense of flavour statics, through
certain constraints on the Yukawa sector of the resulting MSSM,
or, in other words, through the testable predictions for the fermion masses
and mixing angles.
\vspace{5mm}

{\bf 2. Deducing the Mass Matrix: Inverse Hierarchy and Seesaw }
\vspace{3mm}

The mass spectrum of the quarks and charged leptons is spread over five orders
of magnitude, from MeVs to 100 GeVs.
In order to understand its shape it is
necessary to compare the fermion running masses at the scale $\mu\sim M_G$,
where the relevant new physics could take the place. In what follows,
with an obvious notation we indicate by $u,d,...$ the fermion running masses
at the GUT scale, and by $m_u,m_d,...$ their physical masses. For the light
quarks (u,d,s) the latter traditionally are taken as running masses at $\mu=1$
GeV \cite{Leut}.  In doing so, we see that the {\em horizontal}
hierarchy of quark masses exhibits the approximate scaling low (see Fig. 1)
\be{qh}
t:c:u\sim 1:\eps_u:\eps_u^2\,,~~~~~~~
b:s:d\sim 1:\eps_d:\eps_d^2\,
\ee
where $\eps_u^{-1}=200-300$ and $\eps_d^{-1}=20-30$. As for the charged
leptons, they have a mixed behaviour:
\be{lh}
\tau :\mu : e\sim 1:\eps_d:\eps_u\eps_d
\ee
One can also observe that the {\em vertical} mass splitting is small within
the first family of quarks and is quickly growing with the family number:
\be{vert}
\frac{u}{d}\sim 1\,,~~~~\frac{c}{s}\sim 10\,,~~~~
\frac{t}{b}\sim 10^2\,,
\ee
whereas the splitting between the charged leptons and down quarks (at large
$\mu$) remains considerably smaller -- the third family is almost unsplit:
\be{btau}
b\approx \tau \,,
\ee
whereas the first two families are split by a factor about 3 but
\be{ds}
d s\approx e \mu\,.
\ee
One can also exploit experimental information on the quark mixing.
The weak transitions dominantly occur inside the families, and are suppressed
between different families by the small Cabibbo-Kobayashi-Maskawa
(CKM) angles \cite{Nir}:
\be{CKM}
s_{12}\sim \sqrt{\eps_d}\,,~~~~ s_{23}\sim \eps_d\,,~~~~
s_{13}\sim \eps_d^{2}\,,
\ee
This shows that the quark mass spectrum and weak mixing pattern are strongly
correlated. Moreover, there are intriguing relations between
masses and mixing angles, such as the
well-known formula $\,s_{12}=\sqrt{d/s}\,$ for the Cabibbo angle.
All of the observed CP-violating phenomena can be successfully
described in the frames of the Standard Model with the mixing angles in the
range of Eq. (\ref{ds}) and the sufficiently large ($\delta\sim 1$)
CP-phase in the CKM matrix.


It is tempting to think that the certain structure
of the mass matrices is responsible for several mass relations
and the CKM angles can be expressed (explicitly or implicitly)
as functions of the fermion masses.
It is natural to think that these functions have the following
`analytic' properties \cite{decoupling}:

\vspace{1mm}
\underline{\em  Decoupling}: The mixing angles of the first quark family
with others, $s_{12}$, $s_{13}$, vanish in the limit $u,d\rightarrow 0$.
At the next step, when $c,s\rightarrow 0$, $s_{23}$ also vanishes.

\vspace{1mm}
\underline{\em Scaling}: In the limit when the masses of the up and
down quarks are proportional to each other: $u:c:t=d:s:b$, all mixing
angles: $s_{12}$, $s_{13}$ and $s_{23}$ are vanishing.
\vspace{1mm}

Motivated by observations made above,
let us consider for the mass matrices at the GUT scale the following
ansatz:
\be{mf}
\hat{m}_{f}= \mu_f
(\hat{Q}_3+\eps_f\hat{Q}_2+\eps_f^2\hat{Q}_1 )\,,~~~~~~~f=u,d,e
\ee
where $\mu_{u,d,e}$ are some `starting' mass parameters, essentially
the masses of the third family fermions: top, bottom and tau.
The small expansion parameters $\eps_f=\eps_u,\eps_d,\eps_e$ are also
different for the upper quark, down quark and lepton mass matrices.
$\hat{Q}_{1,2,3}$ are some rank-1 matrices with $O(1)$ elements,
not orthogonal in general. These are axis fixing the same 'sceleton' for all
mass matrices $\hat{m}_f$.
Without loss of generality one can take
\be{Q}
\hat{Q}_1\!=\!(0,0,1)^T\!\!\bullet(0,0,1), ~~~
\hat{Q}_2\!=\!(0,a,b)^T\!\!\bullet(0,a'\!,b'\!), ~~~
\hat{Q}_3\!=\!(x,y,z)^T\!\!\bullet(x'\!,y'\!,z')
\ee
so that the mass matrices of Eq. (\ref{mf}) have the form
\be{form}
\hat{m}_f\propto\,\mu_f
\matr{O(\eps_f^2)}{O(\eps_f^2)}{O(\eps_f^2)}
{O(\eps_f^2)}{O(\eps_f)}{O(\eps_f)}
{O(\eps_f^2)}{O(\eps_f)}{O(1)}\,,~~~~~~~f=u,d,e
\ee
In lowest order their eigenvalues are given by diagonal entries and
their ratios are $O(\eps^2):O(\eps):O(1)$, which reproduces the quark
mass pattern of Eq. (\ref{qh}). On the other hand, since $\eps_u\ll\eps_d$,
the quark mixing arises dominantly due to diagonalization of the down quark
mass matrix $\hat{m}_d$ -- the up quark matrix $\hat{m}_u$ is much more
"stretched" and essentially
close to its diagonal form, so that it brings only $O(\eps_u/\eps_d)$
corrections to the CKM mixing angles.
Thus, we expect that $s_{12},s_{23}\sim \eps_d$ and
$s_{13}\sim\eps_d^2$.
Clearly, both the {\em decoupling} and {\em scaling} hypothesis
are naturally fulfilled. In the limit when $\eps_u=\eps_d$
we have the {\em scaling}: $u:c:t=d:s:b$ and all CKM angles are vanishing.
The {\em decoupling} can be seen in the following way: by putting
$\eps^2_{u,d}$ to zero (as parametrically smaller values compared to
$\eps_{u,d}$), we see that $u,d\rightarrow 0$ and at the same time
$s_{12},s_{13}\rightarrow 0$.
At the next step, by putting $\eps_{u,d}$ to zero, we
see that $c,s\rightarrow 0$ and also $s_{23}\rightarrow 0$.

The mass matrices presented in a manner of Eq. (\ref{form}) imply that
the mass generation starts from the heaviest third family and propagates
to the lighter ones, as it is commonly accepted in a mass matrices
model building. However, this situation, especially in the context of the
$SO(10)$ model, looks somewhat paradoxal. If the third family masses
appear from the tree level Yukawa couplings, then only Higgs 10-plet is
good (among the possible 10, 120 and 126) for understanding the $b-\tau$
unification: $\mu_d\approx \mu_e$.
On the other hand, the top and bottom masses are
extremely strongly split: $\mu_u\gg \mu_d $.
These can be reconciled at a price of extremely large tan$\beta$, of about
two orders of magnitude.
However, then it is difficult to understand why are the other parameters in
mass matrices adjusted so that $c/s$ splitting becomes order of magnitude
less compared to $t/b$, and $u$ and $d$ are almost unsplit versus giant
tan$\beta$.
The pattern of vertical splitting (\ref{vert}) suggests that the `large' mass
parameters $\mu_f$ are linked with the expansion parameters $\eps_f$ by
relations $\mu_f\approx m/\eps_f^2$, where the `small' mass parameter
$m$ is the same for all $f=u,d,e$. The structure which appears in this
case can be easier seen by inverting the mass matrices of Eq. (\ref{mf}):
\be{mf-1}
\hat{m}_{f}^{-1}= \frac{1}{m}
(\hat{P}_1+\eps_f\hat{P}_2+\eps_f^2\hat{P}_3 )\,,~~~~~~~f=u,d,e
\ee
where $m\simeq u,d,e$ is a mass scale of the first family fermions. The
rank-1 matrices $\hat{P}_{1,2,3}$ can be taken so that
the inverted mass matrices have the form
\be{form-1}
\hat{m}_f^{-1}\propto\,\frac{1}{m}
\matr{O(1)}{O(\eps_f)}{O(\eps_f^2)}
{O(\eps_f)}{O(\eps_f)}{O(\eps_f^2)}
{O(\eps_f^2)}{O(\eps_f^2)}{O(\eps_f^2)}\,,~~~~~~~
\eps_f=\eps_u,\,\eps_d,\,\eps_e
\ee
In this way, quark mass pattern of Eqs. (\ref{qh}) and (\ref{vert}) is
understood by means of one parameter $\eps_d/\eps_u>1$: we have
$u\!\sim\! d\!\sim \!m$, $c/s\!\sim (\eps_d/\eps_u)>1$ and
$t/b\sim (\eps_d/\eps_u)^2\gg 1$. We call the mass matrix pattern given by
Eq. (\ref{form-1}) the {\em inverse hierarchy pattern}.

The above consideration suggests that the mass generation proceeds from
the lightest family to heavier ones. How one could realize such a situation?

Now it is widely accepted the idea \cite{heavyferm} that the fermion mass
generation is related to higher order
operators induced by the exchange of some hypothetical heavy fermions
in vectorlike representations of $G_{SM}$. One can imagine, that there
are some charged fermion states which are allowed to be superheavy by
the GUT symmetry, or become superheavy after GUT breaking (like the right
handed neutrinos in the context of the $SO(10)$ model).
Let us remark that such fermions automatically appear
in the context of extended GUTs ($E_6$, Georgi's $SU(11)$-type models
etc.) as well as in the context of superstring derived models.
that these heavy fermions participate in the mass
generation phenomenon. Namely, one can consider the the simplest possibility
that the ordinary quark and lepton masses appear through their mixing
with the heavy states, just in analogy with the famous seesaw scenario for
neutrinos \cite{seesaw}. Such a mechanism, named subsequently
{\em universal seesaw mechanism}, was suggested in \cite{uniseesaw}, and
it is indeed capable to provide naturally the inverse hierarchy
pattern.\footnote{The
idea of universal seesaw mechanism was also
explored in a number of papers \cite{uniseesaw2}. The inverse hierarchy,
however, corresponds to the spirit of the original paper \cite{uniseesaw}. }

Let us consider the simple case \cite{uniseesaw,Khlopov}
when along with three families of the standard fermions
$f_i$ ($SU(2)_L$ doublets) and $f^c_i$ (singlets), $i=1,2,3$, we introduce
also three additional families of vectorlike 'quarks' and 'leptons':
$F_i+F^c_i$ (these are weak singlets, like $f^c$'s).
In the context of the Standard Model they are allowed to be
superheavy. However, in is natural to think that their mass generation is
related to some flavour physics beyond the standard model and therefore
their mass matrices $\hat{M}_F$ carry the structure dictated by the
spontaneous breaking pattern of the relevant symmetry
(underlying GUT or horizontal symmetry). One can imagine also that the
same symmetry reasons forbid direct
mass terms of usual fermions $f$ -- i.e. $ff^cH$ couplings with the Standard
Model Higgs $H$,  allowing  instead couplings of the type $\Ga fF^cH$ and
$G f^cFS$, where $S$ is electroweak singlet fragment of some GUT scalar,
and $\Ga$ and $G$ are the correspinding Yukawa matrices.
Then the total mass matrix of $f$ and $F$ type fermions takes the form
\be{matrseesaw}
\begin{array}{cc}
 & {\begin{array}{cc} f^c~~ &~~ F^c \end{array}}\\
M_{tot}~=~\begin{array}{c}
f \\F \end{array}\!\!\! & {\left(\begin{array}{cc}
0 &  \Ga \langle H \rangle \\
G^T \langle S \rangle & \hat{M}_F  \end{array}\right)}\end{array}
\ee
Assuming that $G\langle S \rangle<\hat{M}_F$,
after integrating out the heavy fermion states we receive the light
fermion mass matrix of the following form:
\be{seesaw}
\hat{m}_f= \Ga \hat{M}_F^{-1} G^T \langle S \rangle\langle H \rangle
\ee
It is quite natural to assume that all the Yukawa couplings are $O(1)$,
and all the flavour structure is contained in the `heavy' mass matrices
$\hat{M}_F$.
In other words, the fermion mass hierarchy is initiated in the heavy fermion
sector, while the
usual light fermions are just the spectators of this phenomenon. The inverse
power of $\hat{M}_F$ in Eq. (\ref{seesaw}) is crucial: by means of the seesaw
mixing the heavy fermion mass hierarchy is transfered to ordinary fermions
in an inverted way. This can provide a firm basis to the
inverse hierarchy pattern of Eq. (\ref{form-1}). It is clear,
that the heaviest ones among the heavy fermions are just the partners of the
first standard family, and its small mass splitting can be just a
reflection of the symmetry limit: namely, the heaviests of heavies can be
heavier than the relevant GUT scale,
so that their mass terms obey the isotopic and quark-lepton symmetries,
which are natural subsymmetries of the $SO(10)$.

The inverse hierarchy pattern of the type (\ref{form-1}) was explored in the
context of radiative mass generation scenario, and several intriguing
predictions were obtained for the fermion masses and mixing \cite{Rattazzi}.
It is clear, however,
that the use of a radiative mechanism to ensure the mass hierarchy in a
heavy fermion sector is in obvious contradiction with the idea of low-energy
supersymmetry. Within SUSY scheme one should think of some tree level
mechanism that could generate the masses of heavy fermions by means of the
effective operators of successively increasing dimension, providing thus a
hierarchical structure to their mass matrix.

\vspace{5mm}
{\bf 3. Inverse Hierarchy in SUSY ${\bf SO(10)}$ Model}
\vspace{3mm}

We intend to build a predictive SUSY $SO(10)$ model for the fermion masses,
pursuing the universal seesaw mechanism in order to obtain naturally the
inverse hierarchy pattern.
For this purpose one has to appeal to some symmetry properties.
We assume that there is some extra `family' symmetry
${\cal H}$, that distinguishes the superfields involved into the game.
In the following we will not specify the exact form of ${\cal H}$, describing
only how it should work. We also demand that our model fulfills
the following fundamental conditions:

\vspace{2mm}
{\em A. Unification of the strong, weak and weak hypercharge gauge couplings}
$-$ correct prediction for $\sin^2\!\theta_W$ or $\alpha_s$ at lower energies.

\vspace{1mm}
{\em B. Natural (not fine-tuned) gauge hierarchy and doublet-triplet
splitting} $-$ a couple of Higgs doublets should remain light whereas
their colour triplet partners in GUT supermultiplet should be superheavy.

\vspace{1mm}
{\em C. Sufficiently long-lived proton} $-$ proton lifetime should be above
the current experimental lower bound $\tau_p>10^{32}$ yr.
In particular, we have to make safe the $X,Y$ gauge boson mediated $d=6$
operators, as well as the colour Higgsino mediated $d=5$ operators and
the squark mediated $d=4$ operators.

\vspace{1mm}
{\em D. Natural suppression of the Flavour changing Neutral currents} (FCNC).
\vspace{2mm}

Let us construct such a SUSY $SO(10)\otimes {\cal H}$ model.\footnote{
For more details see ref. \cite{Kazim}. }
We know that three families of quarks and leptons should be arranged within
16-plets $16^f_i$, $i=1,2,3$. Besides them, I exploit three families of
superheavy fermions $16^F_k+\overline{16}^F_k$. All these have certain
transformation properties under ${\cal H}$ symmetry. For the following it is
convenient to describe them in the terms of $SU(4)\otimes SU(2)_L\otimes
SU(2)_R$ decomposition:
\be{16f}
16^f_i=f_i(4,2,1)+f^c_i(\bar{4},1,2)
\ee
\be{16F}
16^F_i={\cal F}_i(4,2,1)+F^c_i(\bar{4},1,2) \,,~~~~~~~~
\overline{16}^F_i={\cal F}^c_i(\bar{4},2,1)+F_i(4,1,2)
\ee
(Notice that ${\cal F}$'s denote the $SU(2)_L$ doublet states and $F$'s are
weak singlets). The vectorlike fermions (\ref{16F}) can have large bare mass
terms or acquire them via couplings to Higgs 45-plets.
For the electroweak symmetry breaking and quark and lepton mass generation
we use a traditional 10-dimensional Higgs supermultiplet
\be{10}
10=\phi(1,2,2)+T(6,1,1)
\ee
For the $SO(10)$ symmetry breaking we promote, as usual,
a set of scalar superfields, consisting of various 54-plets, 45-plets and
$(126+\overline{126})$-plets, which also have certain transformation
properties under ${\cal H}$.
\beqn{scalars}
54=(1,1,1)+(1,3,3)+(20,1,1)+(6,2,2)~~~  \nonumber \\
45=(15,1,1)+(1,3,1)+(1,1,3,)+(6,2,2)~~ \\
126=(10,1,3)+(\overline{10},3,1)+(6,1,1)+(15,2,2) \nonumber
\eeqn
We suggest that all 54-plets have the VEVs of standard configuration
corresponding to the symmetry breaking channel $SO(10)\rightarrow
SU(4)\otimes SU(2)_L\otimes SU(2)_R$. As for the 45-plets, we suggest that
there are three types of them: $45_{BL}$-type fields with VEV on their
(15,1,1) fragment, breaking $SU(4)$ down to $SU(3)_c\otimes
U(1)_{B-L}$; $45_R$-type fields with VEV on the (1,1,3) component, providing
the breaking $SU(2)_R\rightarrow U(1)_R$; and $45_X$-type fields having the
VEVs on both (15,1,1) and (1,1,3) components, and Finally, the 126-plet with
VEV across the (10,1,3) direction completes
the $SO(10)$ breaking down to $G_{SM}=SU(3)_c\otimes SU(2)_L
\otimes U(1)_Y$:\footnote{For some details of the Higgs superpotential analysis
see e.g. refs. \cite{BabuBarr,Rabi}. Let us remark that
`general' form of the VEV $\langle 45_X \rangle$ with $x\neq 1$ can be
obtained if the superpotential includes the terms $45_X^254$ and
$45_X 126 \overline{126}$. As for $45_{BL}$ and $45_R$,
in order to ensure the strict `zeroes' in their VEVs, they should couple only
to 54-plet, but not to 126-plets.}
\be{vevs}
\begin{array}{l}
\langle 54 \rangle = I\otimes \mbox{diag}(1,1,1,-3/2,-3/2)\cdot V_G \\
\langle 45_{BL} \rangle = \sigma\otimes \mbox{diag}(1,1,1,0,0)\cdot V_{BL} \\
\langle 45_R \rangle = \sigma\otimes \mbox{diag}(0,0,0,1,1)\cdot V_R \\
\langle 45_X \rangle = \sigma\otimes \mbox{diag}(1,1,1,x,x)\cdot V_X \\
\langle 126_{(10,1,3)}\rangle = \langle \overline{126}_{(\overline{10},1,3)}
\rangle = v_R
\end{array} ~~~~
I\!=\!\mat{1}{0}{0}{1}\!,~~~\sigma\!=\!\mat{0}{1}{-1}{0}
\ee
Motivated by the coupling crossing phenomenon in MSSM \cite{Amaldi}, we
suggest that $SO(10)\otimes {\cal H}$ is broken down to $G_{SM}$ at once,
by VEVs of Eq. (\ref{vevs}) $\sim M_G$. Below this scale the
theory is just MSSM, with three fermion families ($f_i$) and
one light couple of the Higgs doublets ($\phi=H_1+H_2$). In other words,
from the SUSY scale $M_S$ up to GUT scale $M_G$ we have {\em Grand Desert}.
Many conditions of the series {\em A - D} are immediately fulfilled in this
way: (i) $\sin^2\!\theta_W(\mu)$ and $\alpha_s(\mu)$ are
correctly related at $\mu=M_Z$ (ii) the FCNC are suppressed
provided that the SUSY breaking sector has {\em universal} structure
(let us remind that in Standard model FCNC are naturally suppressed in
both $Z$-boson and Higgs exchanges \cite{FCNC}) (iii)
large unification scale ($M_G\simeq 10^{16}$ GeV) saves the proton from
unacceptably fast decay mediated by $X,Y$ gauge bosons ($d=6$ operators).
Let us note also, that the SUSY $SO(10)$ theory has automatic matter
parity, under which the spinor representations change the sign while the
vector ones stay invariant. (In fact, this gives a natural ground
to refer the spinor representations as fermionic superfields, and the
vector ones as Higgs superfields.)
Provided that non of the $16$-plets has the
VEV, this implies an automatic $R$-parity conservation for the resulting
MSSM. It is well-known, that the proton decaying $d=4$ operators
mediated by squarks are absent in this case.

Before addressing to the fermion mass issues let us first outline the ways
to solve the Doublet-Triplet splitting puzzle and the problem of
dangerous $d=5$ operators.

\vspace{2mm}
{\bf Doublet-Triplet splitting.} We require that the $\phi(1,2,2)$ component
of 10-plet, which consists of the MSSM Higgs doublets $H_1$ and
$H_2$, remains massless in the SUSY limit. On the other hand, the $T(6,1,1)$
fragment, containing colour triplets, should acquire the mass of the order of
$M_G$:  otherwise it would cause unacceptably fast proton decay and
also would affect the unification of the gauge couplings. In order to resolve
this famous problem without fine tuning of the superpotential parameters, one
can address to the Dimopoulos-Wilczek `missing VEV' mechanism utilizing
the Higgs $45_{BL}$-plet \cite{DiWi}. The VEV of $\phi$ arises after the SUSY
breaking and breaks the $SU(2)_L\otimes U(1)_Y$ symmetry:
\be{phi}
\langle\phi\rangle=\mat{v_2}{0}{0}{v_1},~~~~~~~~
(v_1^2+v_2^2)^{1/2}=v=174\,\mbox{GeV}
\ee
where $\mbox{tan}\beta=v_2/v_1$ is the famous up-down VEV ratio in MSSM
\cite{Kane}.

\vspace{2mm}
{\bf d=5 operators} \cite{Sakai}. The universal seesaw can be succesfully
implemented for the needed suppression of these operators in a manner pointed
out in \cite{proton}. (For the analogous mechanism using heavy $144+
\overline{144}-$plets see also \cite{Dvali}.)
One can assume that ${\cal H}$ symmetry does not allow
direct mass terms for light fermions $f$ (i.e. the Yukawa couplings
$16^f16^f 10$), and they are generated through the seesaw mixing with heavy
ones due to the following terms in the superpotential:
\be{WY}
\Gamma_{ik} 10\, 16^f_i 16^F_k + G_{ik} 45_R\, 16^f_i \overline{16}^F_k
\ee
After substituting the large VEVs the whole $9\times 9$ Yukawa matrix
for the fermions of different charges acquires the form
\be{Mtot}
\begin{array}{ccc}
 & {\begin{array}{ccc} \,f^c\,\, & \,\,\,\;F^c & \,\,\,\,\;{\cal F}^c
\end{array}}\\ \vspace{2mm}
W^f_{Yuk}~=~\begin{array}{c}
f \\ F \\ {\cal F}   \end{array}\!\!\!\!\!&{\left(\begin{array}{ccc}
0 & \hat{\Gamma}\phi  & 0 \\ \pm\hat{G}^TV_R  & \hat{M}_F & 0 \\
\hat{\Gamma}^{\dagger}\phi
 & 0 & \hat{M}_{\cal F} \end{array}\right)}
\end{array}
\ee
(each entry of this matrix is $3\times 3$ matrix in itself, $f=u,d,e,\nu$).
The choice of the VEV $\langle 45_R\rangle$ towards the $(1,1,3)$ direction
implies that the $(1,3)$-block of this matrix is vanishing. Therefore, only
the $SU(2)_L$-singlet $F$-type fragments of eq. (\ref{16F}) are important
for the seesaw mass generation, whereas the ${\cal F}$-type ones are
irrelevant. As it was shown in \cite{proton}, this feature is decisive for
the natural suppression of the $d=5$ operators dangerous for the proton.
Indeed, the $f$ and ${\cal F}$ states are unmixed, so
the colour triplets in $T(6,1,1)$ can cause transitions of $f$'s
only into the superheavy ${\cal F}$'s. Therefore, the baryon number violating
$d=5$ operators of the type $[ffff]_F$ ($f=q,l$) -- so called $LLLL$ operators
which bring the dominant contribution to the proton decay after being dressed
by the Winos, are automatically vanishing. As for the $RRRR$ type
operators $[f^cf^cf^cf^c]_F$ ($f^c=u^c,d^c,e^c$), they clearly appear due to
the $f^c-F^c$ mixing and have their usual size. However, they are known to be
much more safe for the proton (see e.g. \cite{Nath} and refs. therein).

\vspace{2mm}
{\bf Flavour Structure.}
The structure obtained in Eq. (\ref{Mtot}) is a good starting point in order
to proceed towards the fermion mass generation.
It tells that the $(2,1)$-blocks of the matrices (\ref{Mtot}) are essentially
the same for the fermions $f$ of all charges -- they differ {\em only}  by the
sign for the up-type and down-type fermions: it is $+G^TV_R$ for
$f=u,\nu$ and $-G^TV_R$ for $f=d,e$. The $(1,2)$ blocks are also the same:
the matrix $\Gamma$ stands for the coupling of up-type and down-type fermions
with the MSSM Higgses $H_2$ and $H_1$, respectively. On the other hand,
as we already noted, the $(1,3)$-block is vanishing, so that the
${\cal F}$-type fermions are irrelevant for the seesaw mass generation. In
what follows, we assume for the simplicity that $G=\Gamma$ (one can check that
this assumption will not change essentially our results). We do not specify
the form of $\Gamma$, suggesting that it is some general complex and
non-degenerated matrix with elements $O(1)$.
Without loss of generality, by suitable redefinition
of the $f$ fermion basis, we always can bring it to a skew-diagonal form:
\be{skew}
\Gamma_{ik}=0\,,~~~~~    \mbox{if}~~~ i<k.
\ee
Therefore,
Assuming $\hat{M}_F\gg \Ga V_R$, the seesaw block-diagonalization of Eq.
(\ref{Mtot}) reduces our theory to the MSSM with the following Yukawa coupling
matrices for the ordinary quarks and leptons:
\be{Yf}
\hat{\la}_f=\Gamma(V_R\hat{M}_F^{-1})\Gamma^T
\ee
Therefore, all the flavour structure is contained in the heavy $F$ fermion
mass matrices and the light fermion ones are $\hat{m}_f\propto\hat{M}_F^{-1}$.
What remains is to obtain the needed hierarchical
pattern for the heavy mass matrices $\hat{M}_F$.

Let us assume that the
${\cal H}$ symmetry allows the bare mass term ($M\gg M_G$) for the first
heavy family $F_1$ and the mass of the $2^{nd}$ one ($F_2$) is generated via
$45_X$:
\be{F12}
M16^F_1\overline{16}^F_1\,+\,g45_X16^F_2\overline{16}^F_2 \,,
\ee
while the $3^{rd}$ family becomes massive through the effective operator
$(45^2_X\!/M\!)16^F_3\overline{16}^F_3$.
In this case the fermion mass hierarchy will be explained due to small
parameter $\eps\!\sim \!V_G/M$.\footnote{For this $M$ should be about
few times $10^{17}$ GeV, which corresponds to superstring scale. In fact,
there is no contradiction in treating our SUSY $SO(10)\otimes {\cal H}$
theory as a superstring derived one. Obviously, since we utilize the
higher dimensional representations of $SO(10)$, such a theory should be
realized at some higher Kac-Moody level. In particular,
$\kappa>5$, if we use the 126-dimensional representation for the symmetry
breaking and neutrino mass generation purposes \cite{Ibanez}.}
However, it is not enough restrictive to use
this operator without specifying to which of the possible $SO(10)$
channels it acts: $45\times 45\rightarrow 1+45+210$.
In order to be less vague, let us introduce the
additional couple $16^F_0+\overline{16}^F_0$ with mass
$M'\!\sim\! M$ and Yukawa couplings $g'16^F_3\overline{16}^F_0 +
g''16^F_0\overline{16}^F_3$.
Then the mass terms of the $F_3$ states appear at the decoupling of the
heavier $F_0$ states, i.e. after the diagonalization of the mass matrix
\be{MF3}
\begin{array}{cc}
 & {\begin{array}{cc} F^c_3\;\;\; &\;\;\; F^c_0 \end{array}}\\
\begin{array}{c} F_3 \\ F_0 \end{array}\!\!\!\!\! &{\left(\begin{array}{cc}
0 & g'\langle 45_X\rangle \\ g''\langle 45_X\rangle & M'\end{array}
\right)}\end{array}
\ee
As a consequence, we obtain the mass matrices $\hat{M}_F$ of the desired form:
\be{MassF}
\hat{M}_F=M(\hat{P}^0_1+\eps_f\hat{P}^0_2+\sigma\eps_f^2\hat{P}^0_3)\,,
{}~~~~~~~~~\begin{array}{c} \hat{P}^0_1=\mbox{diag}(1,0,0) \\
\hat{P}^0_2=\mbox{diag}(0,1,0) \\ \hat{P}^0_3=\mbox{diag}(0,0,1) \end{array}
\ee
where $\sigma \!\sim \!M/M'\!\sim \!1$, since all Yukawa couplings are assumed
to be $O(1)$. What is new, is that the complex
expansion parameters $\eps_f$ ($f=u,d,e,\nu$)
are related due to the VEV pattern (\ref{vevs}) of the $45_X$:
\beqn{eps4}
\eps_d=\eps_1+\eps_2\,,~~~~~~\eps_e=-3\eps_1+\eps_2  \nonumber \\
\eps_u=\eps_1-\eps_2\,,~~~~~~\eps_\nu=-3\eps_1-\eps_2
\eeqn
Hence, only two of these four parameters are independent:
\be{eps2}
\eps_e=-\eps_d-2\eps_u\,,~~~~~~~~ \eps_\nu=2\eps_e+3\eps_u
\ee

As we have already noted, the heavy states $F$ decouple at the scale $V_R$:
below this scale the effective theory is the MSSM with the Yukawa couplings
given by Eq. (\ref{Yf}). Then the ratio $V_R/M$ is given by
$m_e/v\sim\!10^{-5}$. Taking into account that $M/M_G\sim \eps_d^{-1}\sim 30$,
this implies $V_R\sim 10^{13}$ GeV, i.e. some three orders of magnitude below
the GUT scale $M_G$.\footnote{ This is not much of
a problem neither for gauge coupling unification nor for other issues,
provided that the VEV of the $126$-plet $v_R$ is of the order of $M_G$.
Nevertheless, one may find such a small $V_R$ as unappealing.
In this case we can suggest that the $(2,1)$-block of the
"big" matrix (\ref{Mtot}) appears due to the effective operators
$(1/M)(16_f 45_R)(45_R \overline{16}_F)$ rather than the direct
Yukawa couplings of
the eq. (\ref{WY}). These operators can be
obtained by exchange of the
$16+\overline{16}$ states with $\sim M$ masses. Then $(2,1)$-block is
naturally $\sim10^{13}$ GeV when $V_R\sim M_G$, and it is the same for
all types of fermions. }
The seesaw limit $M_F\gg V_R$ is certainly very good for all light states
apart from $t$-quark,
since their Yukawa couplings are much smaller than $1$. However, since
$\lambda_t=O(1)$, we expect the mass of its F-partner
$M_T$ to be of the order of $V_R$ (remember that the Yukawa couplings are
considered to be $O(1)$).
Thus, to evaluate $\lambda_t$, we have to diagonalize the matrix (\ref{Mtot})
without the restriction $\hat{M}_F\gg V_R$. The result is given by
\be{exactseesaw}
\hat{\la}_f \hat{\la}_f^{\dagger} = \Gamma
\left[1\,+\,\hat{M}_F^{\dagger}(\Ga^T\Ga^{\ast}V_R^2)^{-1} \hat{M}_F
\right]^{-1} \Gamma^{\dagger} \,.
\ee
Notice that, when $V_R\gg\hat{M}_F$, this equation gives the obvious
result $\hat{\la}_f=\Gamma $. On the other hand, when
$V_R\ll \hat{M}_F$, it reduces to the Eq. (\ref{Yf}).

\vspace{5mm}
{\bf 4. Mass and Mixing Pattern}
\vspace{3mm}

First it is useful to study $\hat{\la}_f$ in the seesaw limit (\ref{Yf}).
The exact formula (\ref{exactseesaw}) will only be relevant to evaluate
the top quark Yukawa constant $\la_t$.
Once again, the inverse matrices are easier to analyse.
{}From the eqs. (\ref{MassF}) and (\ref{Yf}) we have
\be{Yf-1}
\hat{\la}_{f}^{-1}=\frac{M}{V_R}(\Ga^T)^{-1}
(\hat{P}^0_1+\eps_f\hat{P}^0_2+\eps_f^2\hat{P}^0_3)\Gamma^{-1}=
\frac{1}{\la}(\hat{P}_1+\eps_f\hat{P}_2+\eps_f^2\hat{P}_3)\,,
\ee
where
$\hat{P}_n\propto (\Ga^T)^{-1}\hat{P}^0_n\Gamma^{-1}$ are still rank-1
matrices, but not orthogonal anymore. We can also choose a basis
of eq. (\ref{skew}) and use a relation
$(\Ga^T)^{-1}\hat{P}^0_1\Gamma^{-1}=(\Gamma_{11})^{-2}\hat{P}^0_1$,
to define $\hat{P}_1=\hat{P}^0_1$ and $\la=\Gamma_{11}^2 V_R/M$.
In other words, without loss of generality we can take
\be{Pbasis}
\hat{P}_1\!=\!(1,0,0)^T\!\!\bullet(1,0,0), ~~~
\hat{P}_2\!=\!(a,b,0)^T\!\!\bullet(a,b,0), ~~~
\hat{P}_3\!=\!(x,y,z)^T\!\!\bullet(x,y,z)
\ee
so that the inverse Yukawa matrices can be rewritten as the following:
\be{invform}
\hat{\la}_f^{-1}\,=\frac{1}{\la}
\matr{1+a^2\eps_f+x^2\eps_f^2}{ab\eps_f+xy\eps_f^2}{xz\eps_f^2}
{ab\eps_f+xy\eps_f^2}{b^2\eps_f+y^2\eps_f^2}{yz\eps_f^2}
{xz\eps_f^2}{yz\eps_f^2}{z^2\eps_f^2}
\ee
One can easily see that the
$O(\eps)$ corrections can be neglected in all entries of the matrix
(\ref{invform}) except the 1,1 element. Indeed, the eigenvalues
of the matrices demonstrate the approximate behaviour of Eqs. (\ref{qh}), so
that $\eps_u\ll\eps_d$. Then the CKM mixing angles are essentially determined
by the down quarks Yukawa matrix $\hat{\la}_d$, and if $O(\eps)$
corrections would be negligible for its all entries, we should have
$s_{12},s_{23}\sim \eps_d$ and $s_{13}\sim \eps_d^2$. For $s_{23}$ and
$s_{13}$ this is really the case. For example, from Eq. (\ref{invform})
we have $s_{23}\simeq |yz/b^2|\eps_d$ and $\la_s/\la_b\simeq |z/b|^2\eps_d$.
Then the experimental observation $s_{23}\sim \la_s/\la_b\sim \eps_d$
implies that $z\sim y\sim b$. Thus, the $\eps_d^2y^2$ term leads only to
a few percent correction to the leading term $\eps_d b^2$
in 2,2 element of the matrix (\ref{invform}) and can be safely neglected.
The similar consideration for $s_{13}$ allows one to neglect also $O(\eps^2)$
corrections in 1,1 and 1,2 elements. As for the Cabibbo angle, we have
\be{Cabibbo}
s_{12}\approx \frac{|\eps_dab|}{|1+\eps_da^2|}\approx
\sqrt{\frac{\la_d}{\la_s}|\eps_da^2|}
\ee
Hence, due to experimentally observed relation $s_{12}\approx \sqrt{d/s}$
we have to assume that $|\eps_da^2|\simeq 1$. In turn, this implies that
these $\eps_{d,e}a^2$ terms in the 1,1 element can split the d-quark
and electron masses by the needed amount.

 One may wonder how to achieve
$\eps_da^2\simeq 1$, if the  Yukawa couplings are assumed to be $O(1)$
and $\eps$ is a small parameter: $\eps_d\!\sim\!1/20\!-\!1/30\,$ (see eq.
(\ref{qh})). However, here we still see the advantage of seesaw mechanism:
$a$ is not the coupling constant but rather their ratio,
due to the "sandwiching" between $\Gamma$'s in Eq. (\ref{Yf}).
Thus, it should not come as a surprise if $a^2\!\sim\! 20-30$ due to some
spread in the Yukawa coupling constants,
$a=\Gamma_{21}/\Gamma_{11}\sim\!4\!-\!5$,
while the Yukawa constants themselves are small enough to fulfill the
triviality bound $\Ga^2_{Yuk}/4\pi<1$. On the other hand, the pattern
of the fermion masses and mixing suggests that such a random enhancement
does not happen for other entries in the matrix (\ref{Yf}), so that the
corresponding $O(\eps)$ corrections are negligible.

Thus, in order to split fermion masses within the first family and
to accomodate large ($\sim\!(\eps_d)^{1/2}$) Cabibbo angle,
the matrix (\ref{invform}) must be diagonalized assuming that
$a^2\eps_{d,e}\sim 1$.
Then for the Yukawa eigenvalues at the GUT scale we have
\beqn{eigen}
\frac{\la_u}{\la}=\frac{1}{|1+\eps_u a^2|}\,,~~~~~
\frac{\la_c}{\la}=\frac{|1+\eps_u a^2|}{|\eps_u b^2|}\,,~~~~~
\frac{\tilde{\la}_t}{\la}=\frac{1}{|\eps_u^2 z^2|} \nonumber \\
\frac{\la_d}{\la}=\frac{1}{|1+\eps_d a^2|}\,,~~~~~
\frac{\la_s}{\la}=\frac{|1+\eps_d a^2|}{|\eps_d b^2|}\,,~~~~~
\frac{\la_b}{\la}=\frac{1}{|\eps_d^2 z^2|}    \\
\frac{\la_e}{\la}=\frac{1}{|1+\eps_e a^2|}\,,~~~~~
\frac{\la_{\mu}}{\la}=\frac{|1+\eps_e a^2|}{|\eps_e b^2|}\,,~~~~~
\frac{\la_{\tau}}{\la}=\frac{1}{|\eps_e^2 z^2|} \nonumber
\eeqn
where $\tilde{\la}_t$ is 'would-be' Yukawa coupling given from the
'seesaw' formula (\ref{Yf}). According to Eq. (\ref{exactseesaw}) it
is related to the top quark genuine Yukawa constant through the following
relation:
\be{la-t}
\la_t=\frac{\tilde{\la}_t}{\sqrt{1+(\tilde{\la}_t/\Ga_{33})^2}}<\tilde{\la}_t
\ee
{}From Eqs. (\ref{eigen}) we can immediately obtain the `mass' relations
\be{top}
\left|\frac{\eps_u}{\eps_e}\right|=
\frac{\la_e\la_{\mu}}{\la_u\la_c} \,,~~~~~~
\left|\frac{\eps_u}{\eps_e}\right|^2=\frac{\la_{\tau}}{\tilde{\la}_t}~~~~~
\Longrightarrow ~~~
\frac{\tilde{\la}_t}{\la_{\tau}}=
\left(\frac{\la_u\la_c}{\la_e\la_{\mu}}\right)^2
\ee
\be{bottom}
\left|\frac{\eps_d}{\eps_e}\right|=
\frac{\la_e\la_{\mu}}{\la_d\la_s} \,,~~~~~~
\left|\frac{\eps_d}{\eps_e}\right|^2=\frac{\la_{\tau}}{\la_b}~~~~~
\Longrightarrow ~~~
\frac{\la_b}{\la_{\tau}}=
\left(\frac{\la_d\la_s}{\la_e\la_{\mu}}\right)^2
\ee
These equations are valid at the GUT scale. In order to discuss their
implications for the fermion masses the renormalization group (RG) running
of the coupling constants has to be considered. We have (see e.g. \cite{RG}):
\beqn{RG}
m_u=\la_u\eta_u A_u B_t^3 v\sin \beta\,,~~~~
m_d=\la_d\eta_d A_d v\cos \beta\,,~~~~
m_e=\la_e\eta_e A_e v\cos \beta  \nonumber  \\
m_c=\la_c\eta_c A_u B_t^3 v\sin \beta\,,~~~~
m_s=\la_s\eta_d A_d v\cos \beta\,,~~~~
m_{\mu}=\la_{\mu}\eta_e A_e v\cos\beta \\
m_t=\la_t A_u B_t^6 v\sin \beta\,,~~~~
m_b=\la_b\eta_b A_d B_t v\cos \beta\,,~~~~
m_{\tau}=\la_{\tau}\eta_{\tau} A_e v\cos \beta \nonumber
\eeqn
where the factors $A_f$ account for the running from the scale $M_G$ to the
SUSY breaking scale $M_S$, induced by the gauge couplings, and $B_t$ includes
the running induced by the top quark Yukawa coupling:
\be{B-t}
B_t=\exp\left[-\frac{1}{16\pi^2}\int_{\ln m_t}^{\ln M_G} \la_t^2(\mu)
{\mbox d}(\ln\mu)\right]
\ee
(for definiteness we take $M_S\simeq m_t$). The factors $\eta_f$
encapsulate the QCD+QED running from $m_t$ down to $m_f$ (or to $\mu=1$ GeV
for the light quarks u,d,s).

The Eq. (\ref{top}) shows that the $\eps_u/\eps_d$ ratio is small:
$|\eps_u/\eps_d|<0.15$. Then, by invoking the $SO(10)$ relation (\ref{eps2}),
the Eq. (\ref{bottom}) approximately
gives the 'mass' relationships between the down quarks and leptons:
\be{bottom1}
\sqrt{\frac{\la_b}{\la_{\tau}}}=
\frac{\la_d\la_s}{\la_e\la_\mu}=\left|\frac{\eps_e}{\eps_d}\right|=
\left|1+2\,\frac{\eps_u}{\eps_d}\right|\approx 1
\ee
Thus, we received  the approximate GUT relationships (\ref{btau}) and
(\ref{ds}) between the down quark and lepton masses,
where $O(\eps_u/\eps_d)$ corrections induce about $30\,\%$ uncertainty.
Although the uncertainty is large, this is a remarkable result!!! It shows
that the main `grand' prise, $b-\tau$ unification, is not
completely lost in our inverse hierarchy approach, as was expected
naively (Indeed, $\la_b$ and $\la_\tau$ appear at $O(\eps_{d,e})$ level
and e.g. the factor 2 difference among the coefficients $\eps_{d}$
and $\eps_{e}$
would cause already factor 4 splitting between $b$ and $\tau$,
analogously to the case of up and down quark splitting). However,
the $SO(10)$ group-theoretical relation (\ref{eps2})
ensures that $|\eps_e|\approx |\eps_d|$.
Running the relation (\ref{bottom1}) from the GUT scale down we obtain that
$m_b=5\pm 2$ GeV and $m_dm_s=700-1300$ MeV$^2$.

Moreover, from the equation $\la_d\la_s\approx\la_e\la_{\mu}$ we derive
$\la_d/\la_e\approx 3$, where we have used the current algebra relation
$\la_s/\la_d=m_s/m_d\approx 22$. Since $|\eps_e|\approx |\eps_d|$, the
Eqs. (\ref{eigen}) show that such a splitting is possible  only if
$|\eps_da^2|= 0.5 - 2$. Then Eq. (\ref{Cabibbo}) leads to
$s_{12}\sim \sqrt{d/s} \sim \eps_d^{1/2}$ as compared to what was naively
expected from the ansatz (\ref{form-1}): $s_{12}\sim \eps_d$.
Thus, the mass splitting within the first family implies that the Cabibbo angle
is {\em parametrically} larger as compared to other mixing angles
(which have their
natural size $s_{23}\sim\eps_d$ and $s_{13}\sim \eps_d^2$) and vice versa,
the large Cabibbo angle points that $|\eps_da^2|\sim 1$ and thereby
to the substantial mass splitting between electron and d-quark.
The relation $\eps_d\!\approx\! -\eps_e$ is crucial, since it splits $\la_d$
and $\la_e$ to different sides from $\la\approx\!\la_u$ by about a factor 2.
Then, according to Eq. (\ref{eigen}), the order of magnitude difference
between $\mu/e$ and $s/d$ follows automatically. In this way, owing to the
numerical coincidence $(\la_e/\la_d)^2\!\sim\! \eps_u/\eps_d\sim\!0.1$, we
reproduce the mixed behaviour of leptons (see Eq. (\ref{lh})). The Eqs.
(\ref{eigen}) also imply
\be{u/d}
\frac{\la_u}{\la_d}=|1+\eps_ea^2|\left(\frac{m_em_s}{m_\mu m_d}\right)^{1/2}
= 0.6-0.7
\ee
Thereby, the mass splitting between u- and d-quarks, $m_u/m_d\approx
(\la_u/\la_d)B_t^3\tan\beta$ is $O(1/2)$, if $\tan\beta$ is small and
$\la_t\geq 1$.

Until now, we have given only the qualitative description of the
fermion mass and mixing pattern following from our ansatz.
The detailed analysis leads to more concrete {\em quantitative} results.
We take as an input the masses of leptons, c-quark and b-quark,
the ratio $\zeta=m_s/m_d=\la_s/\la_d$ and the Cabibbo angle
$s_{12}$.  As we know, only two of complex
expansion parameters are independent: $\eps_e=-\eps_d-2\eps_u$. The value
of Cabibbo angle fixes the modulus $|\eps_da^2|$ in terms of $\zeta$.
Using the RG relations (\ref{RG}) we find $\la_e,~\la_\mu,~\la_\tau,~\la_c$
and $\la_b$ in terms of $m_e,~m_\mu,~m_\tau,~m_c,~m_b,~\tan\beta$ and $\la_t$.
Then we can proceed as follows.
{}From the second equations in (\ref{top}) and (\ref{bottom1}) we express
$|\eps_u/\eps_d|$ and $\arg(\eps_u/\eps_d)$ as functions of $\la_t$ and
$\tan\beta$ (for some fixed $\Ga_{33}$). Using Eqs. (\ref{eigen})
and (\ref{bottom}) we obtain
\be{d/e}
\frac{|1-(\eps_d+2\eps_u)a^2|}{|1+\eps_da^2|}=\frac{\la_d}{\la_e}=
\left(\frac{m_{\mu}/m_e}{\zeta}\right)^{1/2}
\left(\frac{m_b A_e\eta_{\tau}}{m_{\tau} A_d \eta_b B_t}\right)^{1/4}
\ee
out of which we can find $\arg(\eps_da^2)$. Therefore, we can
express the complex parameters $\eps_ua^2$ and $\eps_ea^2$, and
thereby $\la_u/\la_e$, in terms of as yet unknown $\tan\beta$ and $\la_t$.
Finally, using the last equation in (\ref{top}) we can find $\tan\beta$ in
terms of $\la_t$. This can compared
with the relation between $\tan\beta$ and $\la_t$ given by fixing
the top quark physical mass $M_t=m_t[1+4\alpha_3(m_t)/3\pi]$.


The results of numerical calculations are shown on Fig. 2.
We have taken $\alpha_3(M_Z)=0.11$, $m_b=4.4$ GeV and $m_c=1.32$ GeV.
For the RG running factors we have used the results of ref. \cite{RG}.
The constant $\Ga_{33}$ is taken to be 3.3, just at the perturbativity border.
The solid curves show the $\tan\beta$ and $\la_t$ correlation resulting
from our `mass' relations for various values of $\zeta=m_s/m_d$.
The dashed curves are isolines for the fixed values of $M_t$.
We see,
that by setting the lower bound on the top mass as $M_t>150$ GeV (160 GeV)
\cite{CDF} we obtain the lower bound $\zeta>19$ ($\zeta>22$) on the
$m_s/m_d$ ratio. The maximal top
mass that we can achieve in our model by taking the extreme value
$\zeta=25$, is about 165 GeV.

We would like to remark that the above mentioned values of $\alpha_3$, $m_b$,
$m_c$ and $\Ga_{33}$ are taken on their extremes, so that for given $\zeta$
the solid curves on Fig. 2 actually mark the upper borders of allowed
regions.\footnote{Let us remind that the condition
$G=\Ga$ on the couplings of eq. (\ref{WY}), which ensures the symmetric form
of the Yukawa matrices in eq. (\ref{invform}), was imposed by hands.
In fact, this condition also leads to the upper
bound  on $M_t$ provided that the `right' Cabibbo angle does not exceed the
physical `left' one which in itself is much above the naively expected
size $O(\eps_d$) \cite{Kazim}.}
The Fig. 3 shows the $\Ga_{33}$ dependence of our results
for $\zeta=22$. We see that the constant $\Ga_{33}$, which sets the
seesaw 'cutoff' (see Eq. (\ref{la-t})) should be quite close to the
perturbativity bound in order to ensure the sufficiently large $M_t$.
For comparison, we have also shown the curve for $\Ga_{33}=6.6$
violating the perturbativity bound. From Figs. 2 and 3
we also see that the preferable values $\la_t$, at which $M_t$ reaches the
maximum, are in the interval 1 -- 2, which corresponds
to the infrared fix-point \cite{inffix}. As for $\tan\beta$, it  varies
from 1.7 ($M_t=165$ GeV) to 1.2 ($M_t=150$ GeV). For such a small
values of $\tan\beta$ the Higgs sector of MSSM can be tested on LEP200
\cite{kun}.

Running Eq. (\ref{d/e}) from GUT scale down to $\mu=1$ GeV,
we obtain
\be{mdms}
m_dm_s=1050~~ \mbox{MeV}^2
\ee
where the uncertainty in $\la_t$ gives only 1-2 $\%$ correction, provided
that $\la_t<3$. Then for fixed value of $\zeta$ we get
\be{md-ms}
m_d=7.0 \cdot(22/\zeta)^{1/2}\,\mbox{MeV}\,,~~~~~~~
m_s=150 \cdot (\zeta/22)^{1/2}\,\mbox{MeV}
\ee
Finally, we can get some information on the u- and d-quark mass splitting.
The Fig. 3 shows various isocurves (dotted) of the mass ratio $\rho=m_u/m_d$
for the fixed ratio $\zeta=22$. We see that the condition $M_t>150$ GeV
allows to $\rho$ to vary between 0.4 and 0.8. For $M_t\approx 160$ GeV
the allowed interval becomes smaller, $\rho=0.5-0.6$.

We also show the $\rho$-isocurves for $\zeta=19$ (Fig. 4a) and $\zeta=25$
(Fig. 4b). From Fig. 4a
we see that the maximal possible top mass $M_t=150$ GeV implies
$\rho\approx 0.4$. Fig. 4b demonstrates that for the allowed region
the values of $\rho$ is $\rho=0.4-1$ (for $M_t=150$ GeV)
and $\rho=0.5-0.9$ (for $M_t=160$ GeV). For $\zeta=25$ these values are
somewhat higher than is allowed by the
current knowledge on the light quark mass ratios \cite{uds}.

Taking all these into account, we can conclude that the preferable choice
for our ansatz corresponds to the parameter region $m_s/m_d=19-22$,
when $M_t=150-160$ GeV ($\tan\beta=1.2-1.4$) and $m_u/m_d=0.4-0.6$.
These values of top mass are somewhat lower as compared to the CDF
result $M_t=174\pm 10\pm 13$ GeV \cite{CDF}.
One can remark, however, that this {\em preliminary} estimate by CDF
is somewhat controversial.
Namely, their $t\bar{t}$ production cross section corresponds to $M_t\approx
150$ GeV rather than 170 GeV. This may hint that the top mass central
value resulting from the combined fit can be shifted towards $160-165$ GeV.

\vspace{5mm}
{\bf 5. Discussion and Outlook}
\vspace{3mm}

One could imagine that our SUSY $SO(10)\otimes {\cal H}$ theory is what
remains from the superstring after compactification. Obviously, such a
theory should be realized at some higher Kac-Moody level, since
we utilize the higher dimensional representations of $SO(10)$: 54, 45 etc.
The fermionic
sector includes 5 zero modes of $16$-plets: $16^f_{1,2,3}$ and $16^F_{2,3}$,
and 2 zero modes of $\overline{16}$-plets: $\overline{16}^F_{2,3}$. We have
also included in game non-zero modes like $16^F_{1}+\overline{16}^F_{1}$,
with masses of the order of the compactification scale $M\sim$ few times
$10^{17}$ GeV. Taking seriously the coupling crossing phenomenon in MSSM,
we suggest that the breaking of $SO(10)\otimes {\cal H}$ symmetry down to
$SU(3)_c\otimes SU(2)_L\otimes U(1)_Y$ occurs at one step, at the scale
$M_G\sim 10^{16}$ GeV. All what remains below is just MSSM with three
quark-leptonic families.

We assumed that the generation of fermion masses occurs due to universal
seesaw mechanism. Once again we would like to stress that in seesaw picture
the ordinary light fermions $f$ are just the spectators of the phenomena
that determine the flavour structure. This structure arises in a sector of
the superheavy $F$ fermions and is transferred to the light ones at their
decoupling. Namely, the heaviest $F$ family $F_1$ is unsplit since it has
$SO(10)\otimes {\cal H}$ invariant mass of the order of $M$. The lighter ones
$F_2$ and $F_3$ get the masses of the order of $M_G$ and $M_G^2/M$
respectively, due to effective operators involving the
Higgs $45_X$ with successively increasing power, and are thereby split.
As a result, the $f$'s
mass matrices, given by a seesaw mixing with the $F$'s, have the
{\em inverse hierarchy} form given by eq. (\ref{mf-1}).

These mass matrices
reproduce the fermion spectrum and mixing pattern in a very economical way.
They differ only
due to different, in general {\em complex} expansion parameters
$\eps_f\sim M_G/M\sim 10^{-1}-10^{-2}$, where $f=u,d,e,\nu$.
These parameters are related
through the $SO(10)$ symmetry properties (see eq. (\ref{eps2})), so only
two of them, say $\eps_d$ and $\eps_u$ are independent. Due to common
mass factor $m$, the first family plays a role of a {\em mass unification
point}, and the $e-u-d$ mass splitting is understood by the same
mechanism that enhances the Cabibbo angle up to the $O(\sqrt{\eps_d})$ value.
Other mixing angles naturally are in the proper range (see eq. (\ref{CKM})).
We have obtained a number of interesting mass formulas, from which it
follows that $m_t\sim 150$ GeV, $m_b\sim 5$ GeV, $m_s\sim 150$ MeV,
$m_d\sim 7$ MeV and $m_u\sim 4$ MeV. The upper limit on the top mass in our
scheme is about 165 GeV. On the other hand, the CDF lower bound $M_t>150$
GeV (or $M_t>160$ GeV) implies the lower bound on the $s/d$ mass ratio:
$\zeta>19$ (or $\zeta>22$).
Using eq. (\ref{eps2}) for $\eps_{\nu}$, we can calculate also the neutrino
Dirac masses. However, in order to provide any predictions for the neutrino
mass and mixing pattern we have to fix also the Yukawa couplings of the
126-plet. We have also the interesting prediction
$\tan\!\beta=1.2-1.7$. This implies interesting phenomenology for the MSSM
Higgs sector which can be tested on the accelerators under construction.


On the other hand, our seesaw pattern
(\ref{Mtot}) leads to the automatic suppression of the
$LLLL$-type $d=5$ operators dangerous for proton decay, and only much weaker
$RRRR$-type ones remain to be effective. This leaves us with the
chance to observe the proton decay at the level of present experimental bound.
It is worth to remark also, that in our scheme we can evaluate the branching
ratios of the different decay modes, since we are able to calculate
all mixing angles, including the ones for the charged leptons\footnote{
In fact, the existing calculations of the proton decay modes (see e.g.
\cite{Nath}) cannot be satisfactory, since they are performed within the
framework of the minimal SUSY $SU(5)$ model with obviously wrong mass
relations $d=e$ and $s=\mu$. }
This can be rather important for testing the inverse hierarchy
scheme, if the proton decay will be observed in the future.

We have not suggested any concrete example of our misterious
symmetry ${\cal H}$ that could support the inverse hierarchy pattern
of the fermion mass matrix, the Dimopoulos-Wilczek ansatz for
natural doublet-triplet splitting and, finally, the needed
VEV pattern. In fact,
${\cal H}$ could contain some set of discrete or abelian (Peccei-Quinn type)
symmetries. Non-abelian horizontal symmetries like $SU(3)$ or $SO(3)$
can be also implemented. Alternatively, one can try to utilize global or
discrete $R$-symmetries. I believe that to find a viable example of
${\cal H}$ symmetry is a rather
cumbersome but feasible task for a smart model-builder.

We find it amusing that the idea of inverse hierarchy, implemented in
SUSY $SO(10)$ theory in a natural way, can explain the key features of the
fermion mass spectrum and weak mixing. Notice, that in contrast to all the
known predictive frameworks for fermion masses (see e.g.
\cite{Fritzsch,DHR}),
we did not exploit any particular `zero' texture
for the Yukawa matrices. Moreover,
it is clear that in our matrices (\ref{invform}) there can be no `zeroes' -
this would immediately lead us to obviously wrong predictions. However, a
{\em clever} horizontal structure in $\Ga$ can reduce the number of input
parameters and thus enhance a predictive power
of our approach. In particular, the ${\cal H}$ symmetry could induce
the matrix $\Gamma$ with certain `zero' texture. These `zeroes' will not be
seen directly in the quark and lepton mass matrices of Eq. (\ref{mf}), but
will manifest themselves through imposing certain relations between the
parameters of Eq. (\ref{Pbasis}). We can also restrict the matrix $\Ga$ to
be real by imposing the CP-ivariance
(spontaneously broken by the complex VEVs in (\ref{vevs})).
This will allow us to calculate also the
CP-violating phase in CKM matrix.

Our approach is an alternative to the other popular predictive SUSY $SO(10)$
frameworks (see e.g. \cite{DHR}), in which the third family
masses appear via direct Yukawa coupling to the Higgs 10-plet, and the
lighter fermion masses are generated by certain higher order operators
with specific $SO(10)$ Clebsch structures. These models have interesting
predictions for fermion masses and mixing angles. However, they seem to be
too much `fine tuned' in order to describe the observed mass pattern.
Indeed, one `fine tuning' should be payed for the extremely large
$\tan\beta$ (of about 60) in order to split top from bottom in spite
of equal Yukawa couplings (see e.g. "The good, the bad and the ugly"
paper by Rattazzi {\em et al.} \cite{tanb}).
Furthermore, the judicious selection of the $SO(10)$ Clebsch structures
is needed to achieve, despite giant $\tan\beta$, the
order of magnitude less splitting between the charm and strange quark masses,
and especially, to obtain  almost unsplit up and down quarks.
These puzzles are completely absent in our `bottom-up' inverted way of
looking on the fermion mass spectrum, which in fact advocates the small
$\tan\beta$ scenario in SUSY $SO(10)$. We assume that the Yukawa
unification takes place for the first family rather than for the third one.
The Yukawa couplings for the second and third families
involve the `up-down' symmetry breaking VEVs in successively increasing
powers, which naturally explains their increasing splitting.
Last but not least, we are
not loosing the understanding of $b-\tau$ unification in spite of naive
expectation. It is more precise the more $t-b$ are split, and
this is granted by the $SO(10)$ symmetry relation (\ref{eps2}).

\vspace{7mm}
{\bf Acknowledgements}
\vspace{2mm}

It is a pleasure to thank Gordy Kane and other organizers of the
SUSY'94 Workshop. I thank Riccardo Barbieri, John Chkareuli, Murman
Margvelashvili and Stephan Pokorski for useful discussions.
Earlier fruitfull discussions and collaboration on related subjects
with Gia Dvali and Riccardo Rattazzi are also gratefully acknowledged.
Special thanks are due to Dr. Anna Rossi for the help in numerical
computations and to Dr. Ursula Miscili for encouragement.

\newpage

\newpage

{\large Figure Captions}

\vspace{7mm}
Fig. 1. Logaritmic plot of fermion running masses at the GUT scale
versus family number. Points corresponding to the fermions with the
same electric charge are joined.

\vspace{3mm}
Fig. 2. Predictions of the model corresponding to different values of
$m_s/m_d$: $\zeta=19$, 22 and 25 for $\Ga_{33}=3.3$ (solid).
Isocurves corresponding to fixed top mass:
$M_t=150$, 160, 170 and 180 (in GeVs) are also shown (dashed).

\vspace{3mm}
Fig. 3. Predictions of the model with different
'seesaw cutoff': $\Ga_{33}=1.5$, 2.2, 3.3 and 6.6, for $\zeta=22$ (solid).
The isocurves corresponding to different values of $m_u/m_d$ are also
shown: $\rho=0.4$, 0.5, 0.6, 0.7 and 0.8 (dotted).

\vspace{3mm}
Fig. 4. Predictions for $\rho=m_u/m_d$ (dotted curves) versus $\zeta=m_s/m_d$:
$\zeta=19$ (Fig. 4a) and $\zeta=25$ (Fig. 4b).

\end{document}